**Critical Shortfall in NIH Support for Medical Physics Research**

Guillem Pratx,[1*] Wensha Yang,[2*] Matthew L Scarpelli[3*]

1. Department of Radiation Oncology, Stanford University, Stanford, CA
2. Department of Radiation Oncology, University of California San Francisco, San Francisco, CA
3. Department of Medical Physics, Purdue University, West Lafayette, IN

* co-corresponding authors

Guillem Pratx, PhD
Associate Professor
Radiation Oncology & Medical Physics
Stanford University
pratx@stanford.edu

Wensha Yang, PhD
Professor
Radiation Oncology & Medical Physics
University of California, San Francisco
Wensha.Yang@ucsf.edu

Matthew L Scarpelli, PhD
Assistant Professor
School of Health Sciences
Purdue University
mscarpel@purdue.edu

Guillem Pratx, Wensha Yang, and Matthew Scarpelli received funding support from NIH. This report is based on volunteer service to AAPM and is not covered by NIH grants.

Acknowledgment: We thank the AAPM Research Committee and the Science Council for their support in reviewing our analysis.

**Abstract**

This report summarizes changes in federal research funding to the medical physics community between FY24 and FY25. By linking the AAPM membership database with NIH RePORTER records, we quantified the distribution of NIH funding for projects led by AAPM researchers. Although total NIH funding to AAPM members remained relatively stable across the two years, the composition of that funding shifted substantially. Competing (new and renewal) awards declined 50%, driven largely by an 80% collapse in new R01 grants from the National Cancer Institute (NCI). In contrast, noncompeting continuation awards increased by 10%, following a shift in how NIH funds multi-year projects. These changes occurred in the context of widespread disruptions to NIH review and grantmaking, including delayed study sections and more stringent administrative requirements. Federal funding is essential to sustaining innovation, supporting early-stage investigators, and ensuring that patients receive the best possible care. The trends identified here raise concerns about the long-term vitality and stability of the medical physics research pipeline.

**Introduction**

Medical physics is a scientific discipline that applies the principles of physics to the diagnosis and treatment of human disease. Medical physicists contribute essential expertise to modern healthcare, supporting technologies ranging from diagnostic imaging and radiation therapy to image-guided procedures and emerging computational and quantitative methods.

Advances in medical physics have been largely driven by decades of federally funded research. In the United States, the National Institutes of Health (NIH) is the primary source of federal research funding for our field. To better understand the relationship between federal investment and research activity in medical physics, the American Association of Physicists in Medicine (AAPM) Working Group for the Development of a Research Database (WGDRD) previously established a database of NIH awards to AAPM members[1]. This database has been updated annually and provides a valuable foundation for analyzing NIH funding trends and the role AAPM members have played in federally supported research[2,3]. In 2022, WGDRD merged with the Working Group for Funding and Grantsmanship (WGFG). The combined group continues to track NIH funding received by AAPM members, guide the selection of timely and relevant topics for the association's annual grant symposia, and answer pertinent questions related to the AAPM workforce[4].

Over the past year, federal research funding processes experienced significant and unprecedented disruptions, including grant cancellations, delays in peer review, and the introduction of new administrative requirements. While these challenges affected many scientific fields, this report specifically quantifies their effects on medical physics research. Using the AAPM NIH Funding Database, we compare funding patterns between FY24 and FY25, analyzing changes across grant mechanisms, NIH institutes, and project activities.

Our analysis reveals several concerning trends, including a pronounced decline in new research awards from the National Cancer Institute (NCI), the primary funder of medical physics research in the United States.

**Methods**

The AAPM NIH Funding Database was constructed using a previously developed algorithm that cross-references NIH RePORTER award records with the AAPM membership database using the principal investigator (PI) name and institution as the primary identifiers for matching. Because the automatic matching can produce false positives, potential matches were manually reviewed, and confirmed false-positive names were stored in an exclusion list[1,3].

NIH grant data were downloaded from RePORTER for FY24 and FY25. Each fiscal year spans October 1 through September 30. The AAPM membership data from July 2025 was used for matching. A total of 360 grants were matched and included an AAPM member listed as the contact PI. These grants were extracted from the four NIH institutes that fund the majority of medical physics research: the National Cancer Institute (NCI), National Institute of Biomedical Imaging and Bioengineering (NIBIB), National Heart, Lung, and Blood Institute (NHLBI), and National Institute of Neurological Disorders and Stroke (NINDS). Collectively, these institutes account for more than 85% of NIH funding to AAPM members.

Funding distributions were analyzed using Pivot Tables, stratified by institute, grant activity code, grant type (new, competing renewal, noncompeting continuation, administrative supplement, or multiyear extension), and fiscal year. Cumulative funding curves were generated by plotting total awarded dollars (including indirect costs) as a function of the Notice of Award (NOA) date.

**Results**

Overall, the total NIH funding for research projects led by AAPM members was comparable in FY24 and FY25. The cumulative funding curves (Fig. 1a) represent grants awarded to AAPM members throughout both fiscal years. The funding amounts include both direct and indirect costs, and the award date is counted from the beginning of each fiscal year (October 1st). Despite a slight slowdown in the second half of FY25, funding distribution accelerated in September, resulting in comparable funding totals across the two fiscal years ($95.9M vs $97.3M).

In FY25, NCI remained the dominant funder of medical physics research (61%), followed by NIBIB (24%), NHLBI (9%), and NINDS (6%). This distribution was consistent with the previous fiscal year (Fig. 1b).

An important distinction exists between competing and noncompeting awards. Competing awards are funded after extensive peer-review by panels of experts and include funding for new projects (type 1) and for renewal of previously funded projects that have concluded (type 2). Noncompeting awards, in contrast, are typically made after internal review by a program officer of the progress of an ongoing project. Noncompeting awards are primarily type 5 (continuation of multiyear projects), but also include administrative supplements

(type 3) and multiyear funding extensions (type 4). Type 7 awards, which are issued when a project is transferred to a new institution, were not included in this analysis.

When looking at grant funding through the lens of competing vs. noncompeting grants, a major divergence emerged (Fig. 1c). Funding for competing awards declined by 51% in FY25, whereas, for noncompeting awards, it increased by 10%. In other words, fewer new projects were funded but more funding was distributed to ongoing projects.

When noncompeting awards were evaluated separately, the cumulative funding curves showed nearly identical trends through the first 11 months of each fiscal year (Fig. 2a). However, in the last month of FY25, a notable uptick in funding can be observed. This discrepancy is due to a small number of projects receiving multiyear extensions (type 4). While these extensions do not change the total amount of funding provided to a project over its lifetime, they result in larger award amounts since multiple years of funding are allocated in a single award.

In sharp contrast, competing awards displayed a widening gap throughout FY25, culminating in a $14.7M shortfall relative to FY24 (Fig. 2b). Noticeable pauses in award activity are apparent around February and again in June/July, likely reflecting disruptions in NIH peer-review and grant decision making[5].

Larger competing awards (R01 and U01) were disproportionately affected, compared to smaller R21 and R03 mechanisms (Fig. 2c). In addition, the largest decreases were at NCI (-64%) and NINDS (-79%), with more moderate reductions at NHLBI (-41%) and NIBIB (-14%). In a surprising reversal, NIBIB has now nearly overtaken NCI as the primary funder of new medical physics research (Fig. 2d).

Given that much medical physics research is in the area of oncology, we focused specifically on competing R01s from NCI to AAPM members. Cumulative funding curves depict a stark divergence between FY24 and FY25, with FY25 exhibiting a near-complete collapse in funding for medical physics-related R01s (Fig. 3a). This gap is a result of the sharp reduction in the number of new R01s, which went from 17 in FY24 to 4 in FY25 (Fig. 3b). Historical grant data from FY22 and FY23 confirms the unprecedented nature of this decline.

For comparison, we examined NCI's entire portfolio of competing R01s across all disciplines over the same period (Fig. 3c and 3d). Whereas FY24 was characterized by steady funding distribution throughout the fiscal year, FY25 showed irregularities, with alternating periods of slower and faster funding, but ultimately reaching a comparable total amount by year's end ($372M vs $387M).

To better understand the disparity between medical physics and the broader cancer research portfolio, we examined a subset of 25 study sections that collectively reviewed roughly 60% of all funded NCI R01s (Table 1). Although the number of R01 awards fluctuated from year to year within individual study sections, grouping them by focus revealed a notable pattern: R01 funding declined by about 18% in clinically or biologically oriented sections, compared with a 34% drop in technology-focused sections.

**Discussion**

Overall, NIH funding to AAPM members was similar in FY24 and FY25. However, the makeup of the funding changed dramatically. New and competing renewal awards declined by 51%, with the steepest drop from NCI, where competing R01 grants to medical physicists fell by 80% from the prior-year level. The overall total was preserved only because noncompeting continuation funding rose 10%, driven largely by multiyear extensions issued in late FY25. In short, fewer new projects were funded, while existing ones were extended to maintain spending levels.

The trends observed in AAPM members' funding closely mirror broader NIH-wide disruptions in FY25. These included a January 2025 threat of a funding freeze and the termination or suspension of > 5,000 grants, disproportionately affecting projects involving Diversity Equity and Inclusion (DEI), environmental justice, or international collaboration[6]. Additional measures, such as a 15% cap on indirect costs, new administrative requirements, and delayed review cycles, led to pauses in awards from February to July FY25, directly contributing to the widening gap between FY24 and FY25 in competing awards.

At the NCI, budget constraints were particularly severe, with a flat $7.22 billion appropriation, mandatory 50% upfront commitments for remaining research project grants and noncompeting awards funded at only 90% of committed levels, which reduces the number of competing awards NCI can fund. Considering the significant budget reductions proposed for FY26, interim FY25 paylines for competing R01s were reduced to the 4th percentile for established and new investigators, a historic low[7], comparing to 10% for FY24[8]. The combination of already limited NCI budget and policy changes drastically prohibited new and renewal awards in medical physics while pre-allocating funding to multiyear (type 4) extensions to meet year-end spending obligations.

Analysis of recent funding trends across NCI study sections reveals a sharper decline in success rates for technology-focused panels compared to those centered on biology and clinical research. While the distinction between "technology" and "biology/clinical" study sections remains somewhat subjective and is largely informed by authors' practical experience with NCI grant submission and review, the pattern is clear: panels emphasizing methodological, computational, or instrumentation advances have experienced more severe funding constriction. This shift indicates that NCI's FY25 priorities are tilting more strongly toward biologically and clinically oriented science. As much of the research conducted by AAPM members falls within the technology-focused domain, it is likely to have been disproportionately affected by this change in direction.

Historically, NIBIB has ranked as the second-largest NIH funding source for AAPM members, trailing only NCI. This strong position reflects the close alignment between NIBIB's mission in imaging, bioengineering, and computational technologies and AAPM members' core expertise in radiation therapy, diagnostic imaging, and dosimetry. Our analysis of FY24 and FY25 data reveals two noteworthy trends: (1) a moderate reduction in the absolute dollar amount of NIBIB competing awards to AAPM members compared with prior years, and (2) a historic shift in FY25, during which a sharp decline in NCI competing awards brought support to medical physics by the two institutes into near-parity. Several multifaceted factors likely contributed to this outcome. Notably, NIBIB maintained more generous paylines than NCI in the past, which may have encouraged a larger proportion of AAPM investigators to route their applications to NIBIB rather than NCI. Although the precise drivers remain to be fully elucidated, the overall decline in total competing dollars awarded to AAPM members, regardless of institute, remains a significant concern for the medical physics research community. More recent developments at NIH center on the implementation of a Unified Funding Strategy, which now requires all 27 Institutes and Centers to abandon rigid paylines. This shift replaces score-based cutoffs with a broader, more holistic set of criteria, thereby leading to additional uncertainty in how grants will be awarded moving forward[9].

The observed decline in competing awards poses a substantial threat to the long-term sustainability of medical physics research. New and renewal grants (primarily R01s) are the primary mechanism for initiating high-impact investigations, including AI-driven imaging and therapy techniques, adaptive radiotherapy, proton/particle beam therapy, and radiopharmaceutical therapy, fields in which AAPM members have historically accounted for 10-55% of NIH-funded projects[3]. The substantial reduction in these awards risks truncating the innovation pipeline and diminishing our members' capacity to generate transformative imaging and therapeutic technologies.  The disruption of NIH funding decisions might put U.S. leadership in medical physics technology at risk. Funding cuts would allow other competing countries to overtake U.S., costing American future leaders, innovation, jobs, and global influence in patient care.

These funding disruptions are likely to disproportionately impact early-stage investigators (ESIs), whose career development is highly vulnerable to funding instability. In contrast to established principal investigators who typically maintain diversified grant portfolios and larger infrastructures capable of absorbing temporary funding loss, most ESIs depend predominantly on their first R01 award to start and establish the lab. Even short delays can quickly stop ongoing experiments and cause skilled researchers to leave. Although NIH extended eligibility for ESI status in response to FY25 delays[10], this administrative adjustment does not offset the loss of research momentum.

Several limitations of our analysis should be noted. First, because inclusion required the contact PI to be an AAPM member, grants involving AAPM members as non-contact PIs may not have been captured. Second, some medical physics researchers are not AAPM members, meaning our analysis represents a subset of the broader research community. Additionally, despite manual efforts to resolve ambiguous PI names, discrepancies may remain due to name variations or shared names across individuals. Finally, because our analysis is based only on funded grants, it does not capture changes in the number of applications submitted and assigned to each institute for review.

In summary, medical physics delivers large impact for patients with remarkably modest resources from NIH. Radiation therapy, enabled by medical physicists, treats ~50% of all cancer patients and contributes to ~40% of cures. Crucial life-saving technologies such as low-dose mammography, image-guided radiation therapy, and advanced magnetic resonance imaging, represent only a tiny fraction of NCI extramural funding, yet they define survival for millions of cancer patients. Ultimately, sustaining research and innovation in medical physics is about ensuring that patients receive the best possible care, a goal that reflects our common dedication to better health for all.

**Conflict of Interest Statement**

The authors are members of the AAPM and receive research funding from the NIH. The content is solely the responsibility of the authors and does not necessarily represent the official views of the NIH or AAPM.

**Figures**

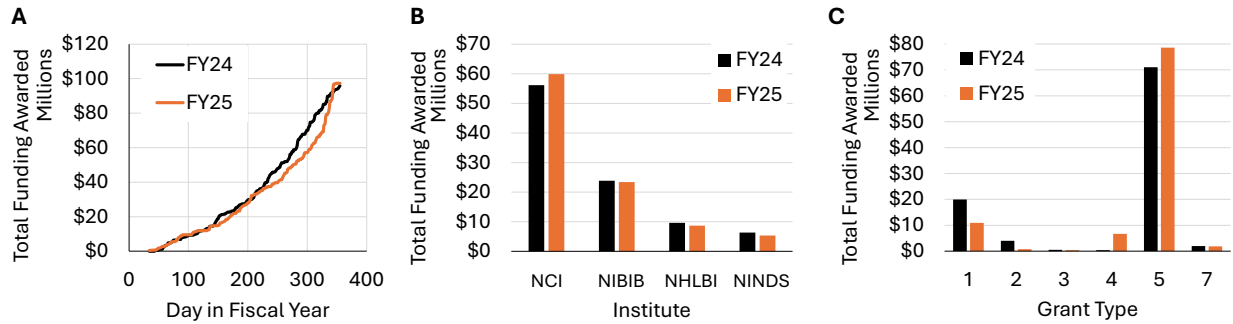

**Figure 1.** Total NIH funding awarded to AAPM members in FY24 and FY25. (a) Cumulative grant funding as a function of award date within each fiscal year. (b) Total funding awarded by NCI, NIBIB, NHLBI, and NINDS, which are top funders of medical physics research. (c) NIH funding for different types of awards, including new projects (1), competing project renewals (2), administrative supplements (3), noncompeting multiyear extensions (4), noncompeting project continuations (5), and institution transfers (7).

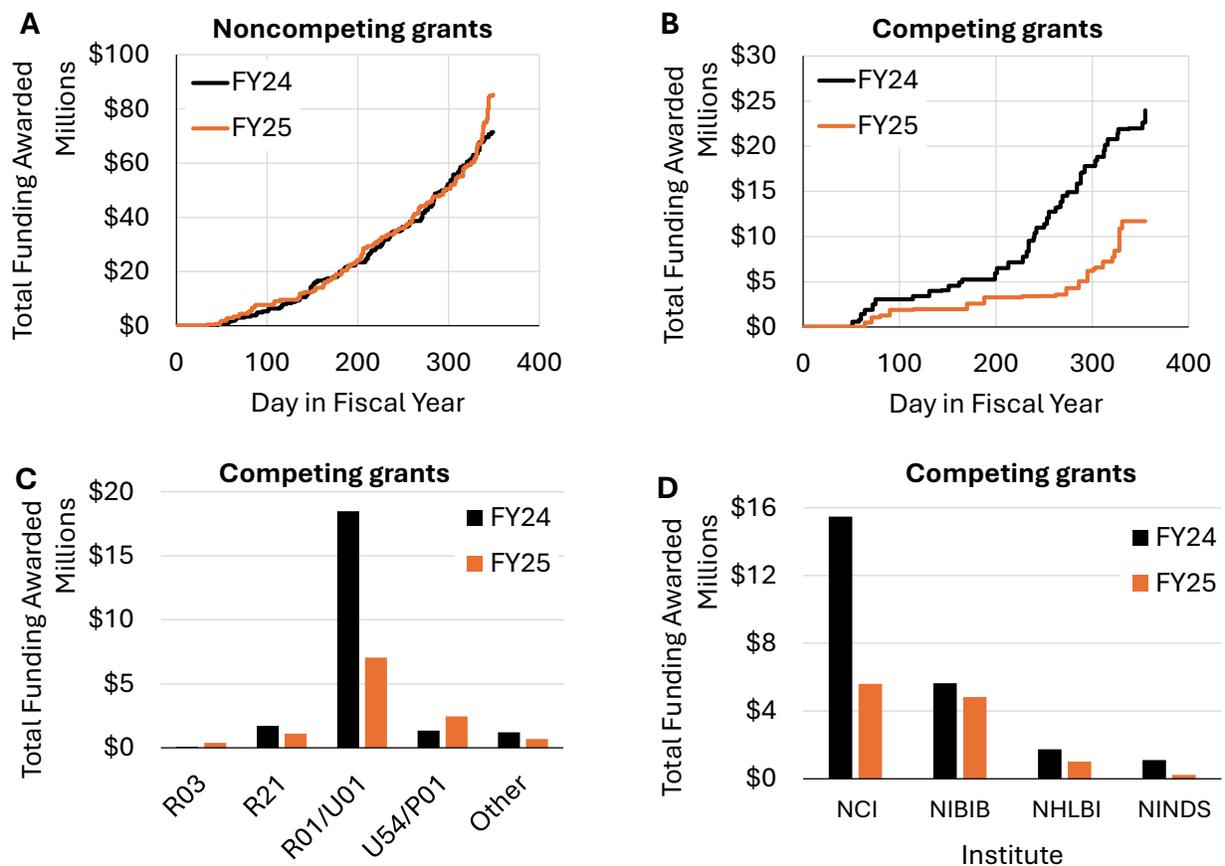

**Figure 2.** Noncompeting and competing awards to AAPM members. (a) Cumulative funding of noncompeting awards (type 3,4 and 5). (b) Cumulative funding of competing awards (type 1 and 2). (c) Competing awards by activity code, including small research grants (R03), exploratory grants (R21), research project grants (R01), cooperative agreements (U01), program project grants (P01) and specialized centers (U54). (d) Funding of competing awards by NCI, NIBIB, NHLBI, and NINDS.

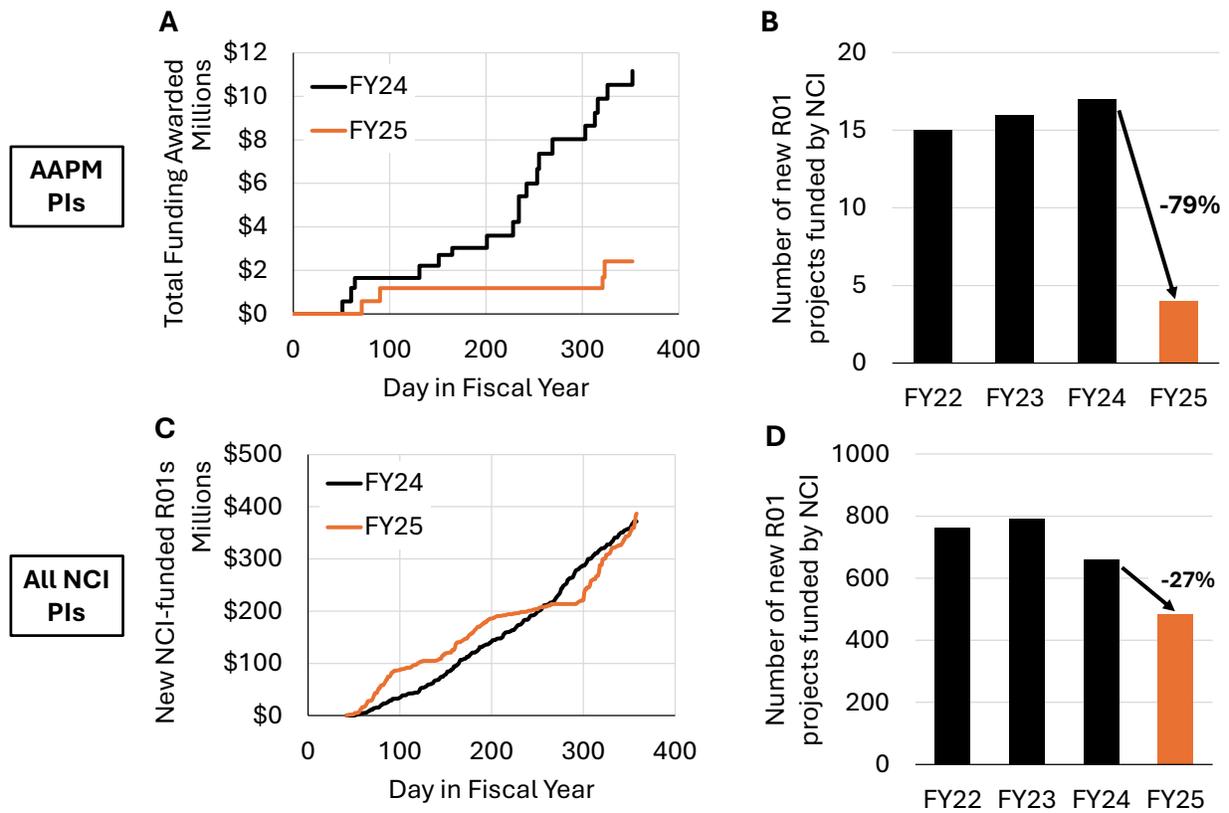

**Figure 3.** NCI funding of competing R01s. (a) Cumulative funding curve for AAPM principal investigators (PIs). (b) Number of competing R01 awarded to AAPM members over the past four years. (c) Cumulative funding allocated by NCI to competing R01s for all investigators. (d) Total number of competing R01s awarded by NCI in the past 4 years.

**Table 1.** Number of R01 projects funded by NCI for a subset of relevant study sections, grouped by general focus (top: biology/clinical; bottom: technology development).

| | Study section | FY24 | FY25 | | |
|---|---|---|---|---|---|
| **Clinical and Biological Research** | Advancing Therapeutics A [ATA] | 26 | 11 | -58% | |
| | Basic Mechanisms of Cancer Health Disparities [BMCD] | 8 | 8 | 0% | |
| | Biochemical and Cellular Oncogenesis [BCO] | 13 | 15 | 15% | |
| | Cancer and Hematologic Disorders [CHD] | 13 | 9 | -31% | |
| | Cancer Cell Biology [CCB] | 17 | 13 | -24% | |
| | Cancer Genetics [CG] | 17 | 12 | -29% | |
| | Cancer Prevention [CPSS] | 9 | 15 | 67% | |
| | Cellular Immunotherapy of Cancer [CIC] | 19 | 13 | -32% | |
| | Clinical Oncology Study Section [CONC] | 17 | 19 | 12% | **-18%** |
| | Drug Discovery and Molecular Pharmacology C [DMPC] | 17 | 18 | 6% | |
| | Gene Regulation in Cancer [GRIC] | 17 | 12 | -29% | |
| | Mechanisms of Cancer Therapeutics A [MCTA] | 27 | 18 | -33% | |
| | Mechanisms of Cancer Therapeutics B [MCTB] | 21 | 20 | -5% | |
| | Mechanisms of Cancer Therapeutics C [MCTC] | 19 | 16 | -16% | |
| | Molecular Cancer Diagnosis and Classification [MCDC] | 19 | 19 | 0% | |
| | Translational Immuno-oncology [TIO] | 11 | 12 | 9% | |
| | Tumor Evolution, Heterogeneity and Metastasis [TEHM] | 16 | 14 | -13% | |
| | Tumor Host Interactions [THI] | 26 | 11 | -58% | |
| **Technology** | Clinical Translational Imaging Science [CTIS] | 5 | 5 | 0% | |
| | Emerging Imaging Technologies and Applications [EITA] | 7 | 5 | -29% | |
| | Imaging Guided Interventions and Surgery [IGIS] | 11 | 5 | -55% | |
| | Imaging Probes and Contrast Agents [IPCA] | 6 | 4 | -33% | **-34%** |
| | Imaging Technology Development [ITD] | 5 | 2 | -60% | |
| | Radiation Therapeutics and Biology [RTB] | 23 | 17 | -26% | |
| | Academic-Industrial Partnership [ZRG1 CTH-E (57)] | 10 | 6 | -40% | |
| | **All study sections** | **661** | **485** | | **-27%** |